\documentclass[12pt]{iopart}
\usepackage{iopams}

\usepackage{natbib}
\usepackage{epsfig}
\usepackage{amssymb}

\newcommand{\bm}[1]{\mbox{\ignorespaces\boldmath$#1$}}
\def \r{\bm{r}}

\def \d{{\rm d}}
\def \K{{\;,}}
\def \PP{{\;.}}

\begin{document}

\title{Slip avalanches in crystal plasticity: scaling of the avalanche cutoff}

\author{Michael Zaiser\footnote[3]{Tel. +44-131-6505671; Fax  +44-131-6513470;
               e-mail M.Zaiser@ed.ac.uk} and Nikos Nikitas}
\address{The University of Edinburgh, Institute for Materials and Processes,
The King's Buildings, Sanderson Building, Edinburgh EH9 3JL, UK}

\begin{abstract}
Plastic deformation of crystals proceeds through a sequence of
intermittent slip avalanches with scale-free (power-law) size
distribution. On macroscopic scales, however, plastic flow is
known to be smooth and homogeneous. In the present letter we use a
recently proposed continuum model of slip avalanches to
systematically investigate the nature of the cut-off which
truncates scale-free behavior at large avalanche sizes. The
dependence of the cut-off on system size, geometry, and driving
mode, but also on intrinsic parameters such as the strain
hardening rate is established. Implications for the observability
of avalanche behavior in microscopic and macroscopic samples are
discussed.
\end{abstract}


\maketitle

\section{Introduction and Background}

It is now well established both experimentally (Weiss and Grasso
1997, Miguel et al. 2001, Dimiduk et. al. 2006, Richeton et al.
2006) and theoretically (Miguel et. al. 2001, Koslowski et. al.
2004, Zaiser and Moretti 2005) that plastic deformation of
crystalline solids proceeds, on microscopic and mesoscopic scales,
through an intermittent series of bursts ('slip avalanches') with
scale-free size distribution. An overview of experimental and
theoretical investigations of this phenomenon has been given
recently by Zaiser (2006). Plastic strain increments $\Delta
\epsilon$ produced by slip avalanches are power-law distributed,
$p(\Delta \epsilon) \propto \Delta \epsilon^{-\kappa}$ where the
exponent $\kappa$ is found to be approximately 1.5 (Miguel et. al.
2001, Zaiser and Moretti 2005, Dimiduk et. al. 2006). Acoustic
emission experiments (Miguel et. al. 2001) indicate scale-free
behavior over more than 8 orders of magnitude without any apparent
cut-off.

This raises an intriguing question: If we consider a deformation
curve (stress vs. strain in case of strain-controlled testing,
stress vs. time in case of creep testing), then this curve will
consist of a sequence of $N$ avalanches such that the total strain
is given by $\epsilon_{\rm tot} = N \langle \Delta \epsilon
\rangle_N$ where $\langle \Delta \epsilon \rangle_N$ is the
average strain increment produced by each of the $N$ events. We
now evaluate the contribution $\Delta \epsilon_{\rm
max}^{(N)}/\epsilon_{\rm tot}$ of the largest event (of size
$\Delta \epsilon_{\rm max}^{(N)}$) to the total deformation. For
$p(\Delta \epsilon) \propto \Delta \epsilon^{-\kappa}$ with $1 <
\kappa < 2$ we obtain the surprising result that in the limit
$N\to\infty$ this ratio tends towards a finite value less than 1.
In other words, whatever the specimen size, the largest avalanche
should always be directly visible on the macroscopic deformation
curves. Experimentally, however, smooth behavior is observed on
macroscopic scales and deformation bursts become apparent only
when the specimen dimensions are reduced down to the micron scale.

Different explanations have been proposed to resolve this paradox.
Zaiser and Moretti (2005) proposed the avalanche size to be
limited by intrinsic hardening. Along similar lines, Richeton et.
al. (2005) argued that the avalanche size in ice polycrystals may
be limited by strong, grain-size dependent kinematic hardening
induced by the strong plastic anisotropy of the material. Another
possible explanation is that the avalanches might exhibit lamellar
geometry (fractal dimension close to 2). Owing to the small volume
involved in the deformation process, such avalanches would produce
only a small macroscopic strain even if their magnitude is limited
by the system size only. This idea is in agreement with
traditional ideas about deformation localization in slip lines and
slip bands (Neuh\"auser 1983), but also with recent experimental
observations which indicate that deformation localizes in lamellar
regions with local strains distributed according to a power law
with exponent close to 1.5 (Schwerdtfeger et. al. 2007). Unlike
the conjecture of Zaiser and Moretti, this proposition predicts an
{\it extrinsic} limit to the avalanche size, in line with the idea
of self-organized criticality.

In the present letter we use the model of Zaiser and Moretti
(2005) to clarify the respective influences of intrinsic
(hardening) and extrinsic (specimen shape and size, driving mode)
parameters on the size of deformation bursts in crystal
plasticity. A short summary of the model is given in Section 2,
together with its extension to include driving by an external
'machine' of variable stiffness. We then use the model to
investigate avalanche size distributions and establish scaling
relations which allow to estimate the size of the largest
avalanches under various experimental conditions.

\section{Description of the Model}

\subsection{Basic structure of the model}

We consider plastic deformation occurring by crystallographic slip
on a single active slip system. Hence, the deformation state is
completely characterized by the scalar shear strain field
$\gamma(\r)$. In the following we assume without loss of
generality that the slip direction corresponds to the $x$
direction of a Cartesian coordinate system and the slip plane is
the $xz$ plane. The driving force for plastic flow is the
externally applied shear stress $\sigma_{xy} =: \tau_{\rm ext}$
acting in this slip system. The evolution of the shear strain is
governed by the equation
\begin{equation}
\frac{1}{\chi}\partial_t \gamma = \tau_{\rm ext} +
\int\Gamma(\r-\r')\gamma (\r') \d \r' + \delta \tau(\r,\gamma)
\label{depin}\K
\end{equation}
which is formally equivalent to the equation of motion of a
depinning elastic manifold. The fluctuating deformation resistance
$\delta \tau(\r,\gamma)$ mimics the formation and dissolution of
jammed dislocation configurations such as dipoles, multipoles, and
(in 3D) dislocation junctions. It is understood as a random field
with short-range correlations, $\langle \delta \tau(\r,\gamma)
\delta \tau(\r',\gamma') = \langle \delta \tau^2 \rangle
f(\gamma-\gamma') \delta(\r - \r')$ where $f$ is a rapidly
decaying function. For details, see Zaiser and Moretti, 2005, and
Zaiser and Aifantis, 2006.

The second term on the right-hand side of Eq. (\ref{depin})
describes long-range interactions mediated by the elasticity of
the medium and characterized by the elastic Green's function
$\Gamma$. In Fourier space, $\Gamma$ does not depend on the
modulus of the wave vector, indicating an effectively infinite
range of the interactions. An explicit calculation of this
function has been given by Zaiser and Moretti (2005), but in the
following we use an alternative formulation based on a dislocation
representation of the stress field. For plane strain deformation
($\gamma = \gamma(x,y)$) with periodic boundary conditions we may
write
\begin{equation}
\frac{1}{\chi}\partial_t \gamma = \tau_{\rm ext} -
\int\tau(\r-\r')\partial_x\gamma(\r') \d^2 r'  + \delta
\tau(\r,\gamma) \label{depinD}\K
\end{equation}
where $\tau(\r)$ is the shear stress field of a straight edge
dislocation of unit strength plus its periodic images. Similarly,
for a general shear strain field $\gamma = \gamma(x,y,z)$ we may
write
\begin{equation}
\frac{1}{\chi}\partial_t \gamma = \tau_{\rm ext} - \int\tau_{\rm
e}(\r-\r')\partial_x\gamma(\r') \d^3 r'  - \int\tau_{\rm
s}(\r-\r')\partial_z\gamma(\r') \d^3 r' + \delta \tau(\r,\gamma)
\label{depinD}\K
\end{equation}
where $\tau_{\rm e}$ and $\tau_{\rm s}$ are the shear stress
fields generated by screw and edge dislocation segments of unit
strength and unit length, plus their periodic images.

\subsection{Numerical Implementation}

We investigate the rate-independent limit $\chi \to \infty$ and
implement an automaton version of the model on a square lattice,
assuming plane strain geometry. The strain field takes discrete
values $\gamma_{ij}$ at the lattice points. The long-range stress
$\int\tau(\r-\r')\partial_x\gamma(\r') \d \r'$ is evaluated as the
sum of the stresses of edge dislocations of strength
$\gamma_{i+1,j} - \gamma_{ij}$ that are placed midway between the
lattice points in the $x$ direction. The stochastic fields $\delta
\tau_{ij}=\delta\tau(\r_{ij},\gamma_{ij})$ are modelled as
independent Ornstein-Uhlenbeck processes located at the respective
lattice points. We allow positive strain increments only
(rate-dependent implementations which allow the strain to locally
decrease have been shown to yield similar results, see Zaiser
2006). In the simulations, the local strain at the lattice point
$(i,j)$ is increased by a fixed amount $\Delta\gamma_{ij} = \Delta
\gamma_0$ whenever the condition
\begin{equation}
\tau_{\rm ext} - \int\tau(\r_{ij} -\r')\partial_x\gamma(\r') \d^2
r' + \delta \tau_{ij}(\gamma_{ij}) > 0 \label{depin1}
\end{equation}
is fulfilled. The strains are updated in parallel, and the new
stresses are evaluated as detailed previously. The new deformation
resistance is evaluated as
\begin{equation}
\delta \tau_{ij}(\gamma_{ij} + \Delta \gamma_{ij}) = \alpha_{ij}
\delta \tau(\gamma_{ij}) + \sqrt{1 - \alpha_{ij}^2} \langle \delta
\tau_{ij}^2 \rangle w_{ij} \K\label{pin}
\end{equation}
where $\alpha_{ij} = \exp(- \Delta \gamma_{ij}/\gamma_{\rm
corr})$, $\gamma_{\rm corr}$ is a correlation strain which
characterizes the 'memory' of the Ornstein-Uhlenbeck process,
$\langle \delta \tau_{ij}^2 \rangle$ is the mean square amplitude
of the pinning field, and the $w_{ij}$ are statistically
independent Gaussian variables of unit variance. For the process
defined by Eq. (\ref{pin}), the two-point correlation function is
$\langle \delta \tau_{ij}(\gamma_{ij})\delta
\tau_{kl}(\gamma'_{kl})\rangle = \langle \delta \tau_{ij}^2
\rangle
\delta_{ik}\delta_{jl}\exp[|\gamma_{ij}-\gamma'_{kl}|/\gamma_{\rm
corr}]$.

\subsection{External driving and strain hardening}

Strain hardening means that the resistance of a material to
deformation increases with increasing strain. In our case, the
deformation resistance is a random field with zero mean. Strain
hardening can be implemented by increasing the amplitude of this
field. To this end, we increase the local amplitude in Eq.
(\ref{pin}) in proportion with the local strain, $\langle \delta
\tau_{ij}^2\rangle = \langle \delta \tau_{0}^2\rangle (1 + K
\gamma_{ij})^2$ where $K$ is a non-dimensional parameter that we
use to adjust the average hardening rate.

In a strain-hardening system, sustained deformation requires a
driving stress that on average increases with strain. In our
simulations, we drive the system in a quasi-static manner, using
either stress control or displacement control with a machine of
finite stiffness. In a stress-controlled simulation we simply
increase the external stress $\tau_{\rm ext}$ from zero in small
increments $\Delta \tau_{\rm ext}$. After each stress increment we
check for all volume elements whether Eq. (\ref{depin1}) is
fulfilled. At the 'unstable' sites where this is the case, we
increase the local strains by $\Delta \gamma_0$ and re-evaluate
the local deformation resistances. After all sites are updated, we
re-evaluate the local stresses everywhere in the system and
perform another update. This is repeated until all sites are
stable (Eq. (\ref{depin1}) is no longer fulfilled for any site).
Then we again increase the external stress, and so on.

In a displacement-controlled test, we impose the total deformation
$\gamma_{\rm ext}$ and evaluate the stress according to
\begin{equation}
\tau_{\rm ext} = M [\gamma_{\rm ext} - \langle \gamma_{ij}
\rangle] \label{machine}\K
\end{equation}
where $M$ is called the 'machine stiffness'. In a quasi-static
simulation we increase $\gamma_{\rm ext}$ (and thus the stress) at
a small rate until at least one site fulfils Eq. (\ref{depin1})
and becomes unstable. We then keep $\gamma_{\rm ext}$ fixed while
we update in parallel the strains at all unstable sites as
previously outlined. However, after each update we not only
re-evaluate the local stresses and deformation resistances, but
also the external stress which, owing to the increase of $\langle
\gamma_{ij} \rangle$ during the update, is bound to decrease
according to Eq. (\ref{machine}). Once all sites have stabilized,
we again increase the imposed strain, and so on. The procedure can
be visualized as 'pulling' the system with a spring of finite
stiffness: As soon as the system yields, the spring is partly
relaxed and the driving force decreases by an amount that is
proportional to the spring stiffness.

\subsection{Relation with physical parameters of dislocation
systems}

The model as stated above applies to the shear deformation of any
disordered material. When the disorder relates to the dislocation
arrangement in a deforming crystal, scaling relations can be used
to relate the model parameters to physical variables of the
system. To this end, we consider a dislocation system of density
$\rho$ and note that the characteristic correlation length of
internal stress fluctuations in such systems is of the order of
one dislocation spacing $\rho^{-1/2}$ (for a detailed discussion,
see Zaiser and Aifantis 2006). The volume occupied by one lattice
site can be identified with the characteristic volume occupied by
a dislocation segment, $V_0 = \rho^{-3/2}$. The correlation strain
$\gamma_{\rm corr}$ corresponds to the strain accomplished locally
when a dislocation segment crosses this elementary volume,
$\gamma_{\rm corr} = b \sqrt{\rho}$ where $b$ is the modulus of
the Burgers vector of the dislocations. This also defines the
natural unit of strain in a dislocation system. The deformation
resistance $\delta \tau$, as all other stresses, scales like $G b
\sqrt{\rho}$ where $G$ is the shear modulus. From the scaling of
stress and strain it follows that the 'natural unit' of the
hardening rate and the machine stiffness is simply the shear
modulus $G$.

In the following all quantities will be given in these 'natural
units'. The size $s$ of a strain burst is defined as the sum of
all strain increments that occur during a period of activity. The
corresponding macroscopic strain is given by $s/XY$ where the
integers $X$ and $Y$ define the size of the lattice in the $x$ and
$y$ directions.

\section{Results}

\begin{figure}[b] \hspace*{-1.5cm}
\centerline{\epsfig{file=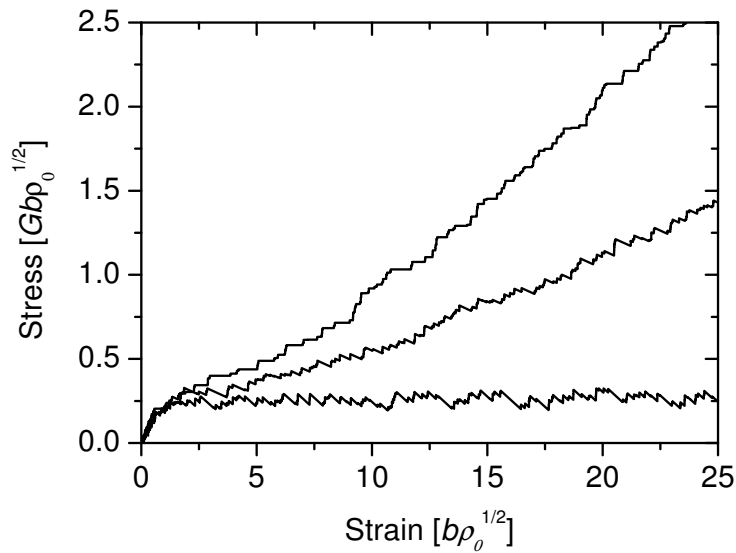,width=10cm,clip=!}}
\caption{Stress-strain curves as obtained from simulation of a
system with 64 $\times$ 64 sites; from top to bottom: hardening
rate $H = 0.125$, machine stiffness $M=0$; hardening rate $H =
0.0625$, machine stiffness $M=0.0625$, hardening rate $H = 0$,
machine stiffness $M=0.125$.}\label{stressstrain}
\end{figure}

As discussed by Zaiser and Moretti (2005), in the absence of
hardening and under conditions of stress control, the model shows
a 'phase transition' between a pinned and a moving phase. This
transition occurs at a critical stress level which defines the
yield stress of the system. In the following we choose the
initial value of $\langle \delta \tau_0^2 \rangle$ such that this
yield stress is about $0.3$ (dimensional value $0.3 G b
\sqrt{\rho})$, in line with typical experimental findings in fcc
metals. The correlation strain is set equal to unity, and the
parameter $K$ is used to tune the hardening rate $H$. We also
systematically vary the machine stiffness $M$ and the system size.
There are no other parameters in the model.

Figure 1 shows three deformation curves (external stress vs.
average strain) obtained for different values of the machine
stiffness and hardening rate. In a stress-controlled simulation
and in the absence of hardening, the external stress passes
through an initial transient until it reaches the yield stress, at
which point the strain increases indefinitely. In a
displacement-controlled simulation, on the other hand, the
external stress fluctuates slightly below the yield-stress level:
Close to the yield stress, slip avalanches are triggered which
decrease the external stress, after an avalanche is terminated the
external stress rises again, and so on. In the presence of
hardening, the mean stress after the initial transient increases
first slowly ('Stage I') and then at a higher, approximately
constant rate ('Stage II'). Statistics of slip avalanches were
determined in this constant-hardening stage (above a strain of
approximately 10 in Figure 1). The hardening rate $H$ was defined
as the average slope of the stress-strain curve in this regime,
with the average either determined from the stress-strain curve of
a very large system or from the average of many smaller systems.
Both methods were found to yield similar results.

\begin{table}[h]
\begin{tabular}{| l | l | l | l |}
\hline
series & fixed parameters & varied parameter & variation range\\
\hline \hline
1 & $M = 1,\; H = 0$ & system size & $32 \times 32 \dots 128 \times 128$\\
\hline
2 & $M = 0.5,\; H = 0$ & system size & $16 \times 16 \dots 128 \times 128$\\
\hline
3 & $M = 0.25,\; H = 0$ & system size & $32 \times 32 \dots 128 \times 128$\\
\hline
4 & $M = 0.125,\; H = 0$ & system size & $64 \times 64 \dots 128 \times 128$\\
\hline
5 & $M = 0,\; H = 0.0065$ & system size & $32 \times 32\dots 128 \times 128$\\
\hline
6 & $M = 0,\; H = 0.013$ & system size & $32 \times 32\dots 128 \times 128$\\
\hline
7 & $M = 0,\; H = 0.026$ & system size & $32 \times 32\dots 256 \times 256$\\
\hline
8 & $M+H = 0.125$, $X = Y = 64$ & stiffness $M$ & 0 \dots 0.125\\
\hline
9 & $M = 0.125$, $H = 0$, $X$ = 16 & extension $Y$ & 16 \dots 128\\
\hline
10 & $M = 0.125$, $H = 0$, $Y$ = 16 & extension $X$ & 16 \dots 128\\
\hline
\end{tabular}
\caption{Series of simulations carried out in this study. Scans
over the variation ranges were performed by changing the
respective parameters by a factor of two between two simulations
in a series.}
\end{table}

\subsection{Truncation of avalanches due to machine-induced stress
relaxation}

\begin{figure}[b] \hspace*{-1.5cm}
\centerline{\epsfig{file=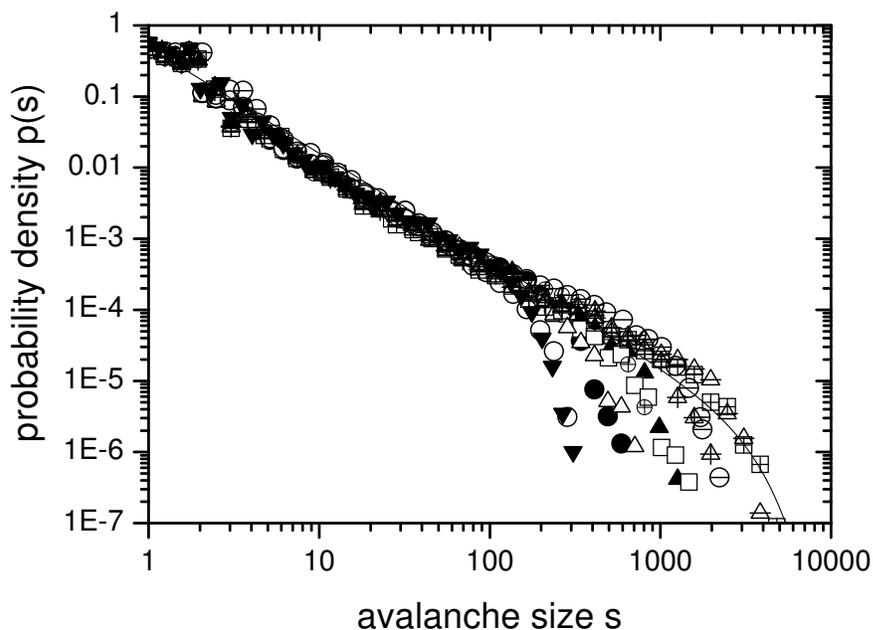,width=12cm,clip=!}}
\caption{Avalanche size distributions corresponding to simulation
series 1 (full symbols), series 2 (open symbols), series 3
(cross-centered symbols) and series 4 (bar-centered symbols).
Symbol shape indicates system size: square $128 \times 128$, up
triangle $64 \times 64$, circle $32 \times 32$, down triangle $16
\times 16$.} \label{avaldist}
\end{figure}

To determine avalanche statistics, several series of simulations
were carried out (Table 1). Generally, the avalanche size
distributions possess the scaling form
\begin{equation}
p(s) \propto s^{-3/2} f(s/s_0)\label{scaledist}
\end{equation}
where $s_0$ is the characteristic avalanche size and the function
$f$ can be well approximated by $f(s/s_0) \approx
\exp[-(s/s_0)^2]$. Distributions corresponding to series 1-4 on
Table 1 are shown in Figure \ref{avaldist}. It can be seen that an
increase in machine stiffness $M$ leads to a decrease and an
increase of system size to an increase in $s_0$. This is to be
expected, since the machine decreases the driving stress, during
an avalanche of size $s$, by an amount $M s/N$ where $N$ is the
number of lattice sites. This causes large avalanches to
self-terminate. A more quantitative analysis can be performed by
fitting Eq. (\ref{scaledist}) to the distributions to determine
the characteristic size $s_0$ as a function of the simulation
parameters. Interestingly, it turns out that $s_0$ scales as $s_0
\propto L/ M $ where $L$ is the linear dimension of the system,
rather than $s_0 \propto N/M \propto L^D/M$ where $D$ is the
system dimension.

\subsection{Equivalence of hardening and machine-induced stress
relaxation}

\begin{figure}[b] \hspace*{-1.5cm}
\centerline{\epsfig{file=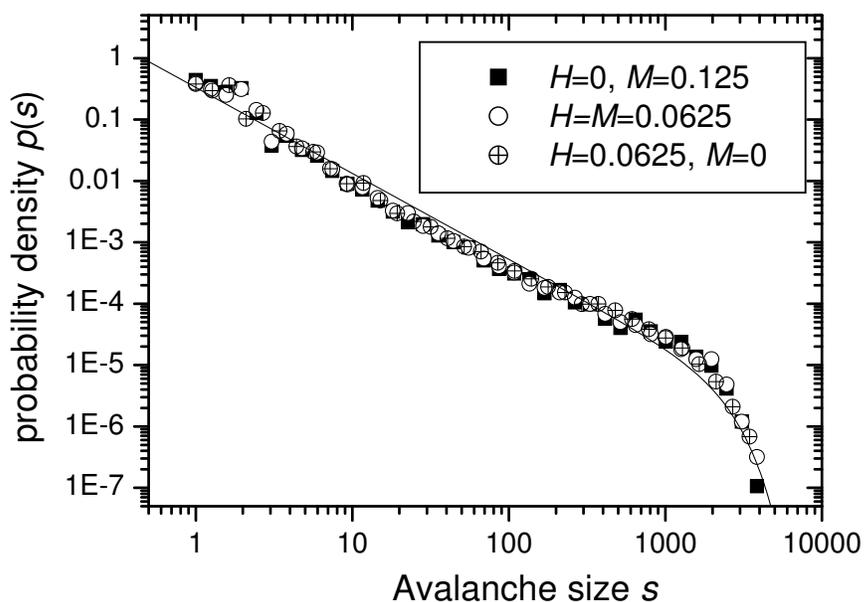,width=12cm,clip=!}}
\caption{Avalanche size distributions for three different systems
of size $64 \times 64$; Parameters $M$ and $H$ see inset; Full
line: fit according to Eq. (\ref{scaledist}), with $s_0 = 1700$. }
\label{machinehard}
\end{figure}

Hardening locally increases the pinning strength of the system at
those sites that are actually involved in a slip avalanche.
Machine-induced stress relaxation decreases the global driving
force on all sites. Both effects should have equivalent
consequences for the avalanche statistics if the depinning
transition of the system shows mean-field behavior.

Figure 3 shows avalanche size distributions determined from an
ensemble of simulations similar to those shown in Figure 1.
Different machine stiffnesses and hardening rates were imposed
such that the sum $M + H$ was the same for all simulations. It can
be seen that three sets of simulations ($M=0, H=0.125\;;
M=H=0.0625\;; M=0.125, H=0$) yield exactly the same distribution
of avalanche sizes, which is well described by the scaling form
(\ref{scaledist}). This demonstrates that strain hardening and
machine-induced stress relaxation indeed have similar effects on
the avalanche dynamics. As a consequence, the avalanche cutoff
$s_0$ depends on the sum $H + M$ of the hardening rate and machine
stiffness, rather than on $H$ and $M$ separately.

\subsection{Scaling law for the characteristic avalanche size}

After performing several series of simulations at different
hardening rates (series 5-7 in Table 1) and determining the
characteristic size $s_0$ for each set of parameters, it was found
that the scaling of the cut-off can, over the entire parameter
range investigated, be described by the general scaling law
\begin{equation}
s_0 = \frac{CL}{(M + H)}\PP
\end{equation}
A fit to all data yields $C \approx 5$ (Figure 4). We also
performed simulations for anisotropic specimen shapes. In this
case we found approximately the same scaling law, with $L$ defined
as the square root of the number of lattice sites (Series 9 and 10
in Figure 4).

\begin{figure}[b] \hspace*{-1.5cm}
\centerline{\epsfig{file=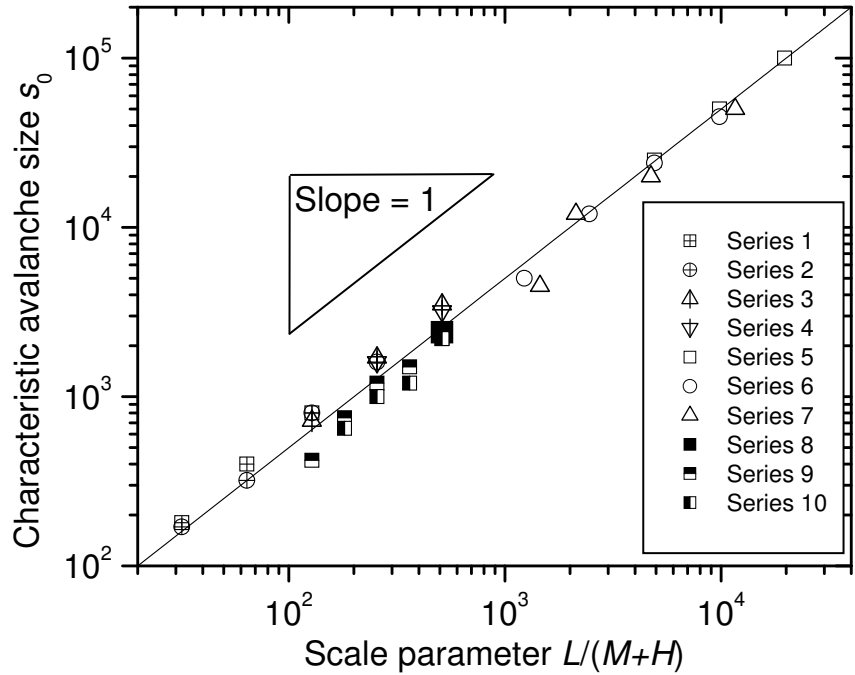,width=12cm,clip=!}}
\caption{Scaling of the avalanche cutoff with hardening rate $H$,
machine stiffness $M$, and linear system dimension $L$. Full line:
$s_0 = 5 L/(M + H)$.} \label{scale}
\end{figure}.

\section{Discussion and Conclusions}

Our results have been obtained in the limit of quasi-static
driving, i.e., the applied stress or the imposed displacement are
kept fixed during an avalanche. This is the relevant limit for
dislocation systems. Direct observation of deformation by surface
monitoring indicates an extremely high level of intermittency in
dislocation plasticity, in the sense that a given small volume
element is plastically inactive almost all of the time: the
instantaneously active slip volume is of the order of $10^{-7}$ of
the macroscopic crystal volume, see Neuh\"auser (1984) and Zaiser
(2000).

To assess the implications of our results for such 'real' systems,
it is useful to revert to dimensional coordinates. The average
strain increment produced by an avalanche of size $s_0$ is given
by $s_0/N$. In dimensional units, this becomes $\Delta \gamma_{\rm
max} = C b G/l(H+M)$ where $l$ is the (dimensional) specimen
length. It is even more instructive to evaluate the length
increment $\Delta l_{\rm max}$ produced by such an avalanche
during axial loading. Using that the axial plastic strain
$\epsilon$ is related to the shear strain by $\epsilon = \Delta
l/l = m \gamma$ where $m$ is the Schmid factor, we find that
\begin{equation}
\Delta l_{\rm max} = \frac{C b G m}{(M + H)}\PP
\end{equation}
This compares well with experimental data on micron-sized single
crystals. Dimiduk et. al. (2006) determined the statistics of
elongation bursts observed during compression of Ni microcrystals.
Specimens with sizes of the order of 20 micrometers were tested
under conditions of load control (the load was kept constant
during a burst). Typical hardening rates found in these
experiments were of the order of $\d\tau_{\rm ext}/\d\langle
\gamma \rangle \approx G/1500$. With $C = 5$, $b = 2.5 \times
10^{-10}$ m, and $m \approx 0.4$, we find that the length
increments caused by the largest bursts are expected to be about
0.7 microns. This compares very well with the maximum burst sizes
observed by Dimiduk et. al. More importantly, the simple scaling
relation predicted in the present work can be easily verified by
comparing specimens of different sizes (the largest elongation
bursts should be roughly independent on specimen size), and by
comparing results obtained in stress and displacement control.

The present results offer a simple explanation of the fact that
avalanches are not observed in macroscopic deformation. Length
increments of the order of less than a micron cannot be resolved
by standard macroscopic testing equipment. In macroscopic samples
with sizes of the order of $10^{-1}$ m, such length increments
correspond to axial strain increments $\Delta \epsilon$ of less
than $10^{-5}$. The proposed scaling relation is consistent with
the idea of a lamellar, system-spanning 'shape' of the largest
avalanches: The prediction that the average strain produced by
these avalanches decreases in proportion with the linear dimension
of the specimen follows naturally from the assumption that each
avalanche corresponds to the formation of a lamellar slip line or
slip band. However, effects of machine stiffness and/or intrinsic
hardening are relevant in delimiting the avalanche size by
controlling the amount of strain produced within such an active
region.

\section*{Acknowledgements}
Financial support of the European Commission under
NEST-2005-PATH-COM-043386 and of EPSRC under Grant No. EP/E029825
is gratefully acknowledged.

\section*{References}
\noindent Dimiduk, D.M., Woodward, C., LeSar, R., Uchic, M.D.,
2006. Scale-Free Intermittent Flow in Crystal Plasticity, Science
26, 1188-1190.\\
\\
\noindent Koslowski, M., LeSar, R., Thomson, R., 2004. Avalanches
and scaling in plastic deformation, Phys. Rev. Letters 93,
125502.\\
\\
\noindent Miguel, M.C., Vespignani, A., Zapperi, S., Weiss, J.,
Grasso, J.-R., 2001. Intermittent dislocation flow in viscoplastic
deformation. Nature 410, 667-671.\\
\\
\noindent Neuh\"auser, H., 1984. Slip-line formation and
collective dislocation motion. In: F.R.N. Nabarro (Ed.),
Dislocations in Solids, Vol. 4, North-Holland, Amsterdam, pp.
319-440.\\
\\
\noindent Richeton, T., Weiss, J., and Louchet, F., 2005.
Breakdown of avalanche critical behaviour in polycrystalline
plasticity. Nature Materials 4, 465-469.\\
\\
\noindent Richeton, T., Dobron, P., Chmelik, F., Weiss, J., and
Louchet, F., 2006. On the critical character of plasticity in
metallic single crystals. Mater. Sci. Engng. A 424, 190-195.\\
\\
\noindent Schwerdtfeger, J., Nadgorny, E.M., Madani-Grasset, F.,
Koutsos, V., Blackford, J.R., Zaiser, M., 2007. Scale-free
statistics of plasticity-induced surface steps on KCl single
crystals. J. Stat. Mech., submitted.\\
\\
\noindent Weiss, J., Grasso, J.-R., 1997. Acoustic emission in
Single Crystals of Ice. J. Phys. Chem. B 101, 6113-6117.\\
\\
\noindent Zaiser, M., Moretti, P., 2005. Fluctuation phenomena in
crystal plasticity - a continuum model. J. Stat. Mech, P08004.\\
\\
\noindent Zaiser, M., Aifantis, E.C., 2006. Randomness and slip
avalanches in gradient plasticity. Int. J. Plasticity 22,
1432-1455.\\
\\
\noindent
Zaiser, M., 2000. Statistical modelling of dislocation
systems, Materials Science and Engineering A 309/310, 304-315\\
\\
\noindent
Zaiser, M., 2006. Scale invariance in plastic flow of
crystalline solids. Adv. Physics 55, 185-245.
\\

\end{document}